\documentclass[aps,pra,twocolumn,superscriptaddress]{revtex4-1}
\usepackage{amssymb}
\usepackage{amsmath}
\usepackage{graphicx}
\usepackage{epsfig}

\begin{document}

\title{Partial topological Zak phase and dynamical confinement in
non-Hermitian bipartite system}
\author{X. Z. Zhang}
\affiliation{College of Physics and Materials Science, Tianjin Normal University, Tianjin
300387, China}
\author{Z. Song}
\email{songtc@nankai.edu.cn}
\affiliation{School of Physics, Nankai University, Tianjin 300071, China}

\begin{abstract}
Unlike a Chern number in $2$D and $3$D topological system, Zak phase takes a
subtle role to characterize the topological phase in $1$D. On the one hand,
it is not a gauge invariant, on the other hand, the Zak phase difference
between two quantum phases can be used to identify the topological phase
transitions. A non-Hermitian system may inherit some characters of a
Hermitian system, such as entirely real spectrum, unitary evolution,
topological energy band, etc. In this paper, we study the influence of
non-Hermitian term on the Zak phase for a class of non-Hermitian systems. We
show exactly that the real part of the Zak phase remains unchanged in a
bipartite lattice. In a concrete example, $1$D Su-Schrieffer-Heeger (SSH)
model, we find that the real part of Zak phase can be obtained by an
adiabatic process. To demonstrate this finding, we investigate a scattering
problem for a time-dependent scattering center, which is a
magnetic-flux-driven non-Hermitian SSH ring. Owing to the nature of the Zak
phase, the intriguing features of this design are the wave-vector
independence and allow two distinct behaviors, perfect transmission or
confinement, depending on the timing of a flux impulse threading the ring.
When the flux is added during a wavepacket travelling within the ring, the
wavepacket is confined in the scatter partially. Otherwise, it exhibits
perfect transmission through the scatter. Our finding extends the
understanding and broaden the possible application of geometric phase in a
non-Hermitian system.
\end{abstract}

\maketitle

%\pacs{11.30.Er, Berry phase, wave packet dynamics?????}
%\author{}

%\begin{center}
%{\Large Partial topological Zak phase and dynamical confinement in
%non-Hermitian bipartite system}
%\end{center}

\section{Introduction}

\label{sec_intro} The scope of quantum mechanics has been extended to
non-Hermitian system since the discovery that a certain class of
non-Hermitian Hamiltonians could exhibit the entirely real spectra \cite%
{Bender1,Bender2,Bender3} and the observation of non-Hermitian behavior in
experiment \cite%
{OLGanainy,PRLMakris,PRLMusslimani,Guo,ruter,Ruschhaupt,Klaiman,Kottos,LonghiLPR}%
. Besides the exceptional point (EP), biorthonormal inner product can be
induced to take the role of Dirac inner product for a pseudo-Hermitian
Hamiltonian operator \cite{Ali,AliPRA}, which always associates with a
particular symmetry, $\mathcal{PT}$ symmetry. Here $\mathcal{P}$\ is an
unitary operator, while $\mathcal{T}$\ is an anti-unitary operator.
Especially, in the $\mathcal{PT}$ symmetric region, a non-Hermitian
Hamiltonian acts as a Hermitian one, having entirely real spectrum, unitary
evolution, etc, in the context of biorthonormal inner product. In this
sense, many conclusions for Hermitian system can be extended to the
non-Hermitian regime. Recently there has been a growing interest in
topological properties of non-Hermitian Hamiltonians applicable to a wide
range of systems including systems with unbalanced pairing, systems with
gain and/or loss, and systems with open boundaries \cite%
{Rudner,Esaki,Malzard,Yuce1,Yuce2,Harter,tonyPRL,Leykam,Weimann,Yin,Lieu,LC,MartinezPRB,Shen,Kunst,Martinez,Xiong,WR,Yao}%
.

In the Hermitian regime, it is well known that the nontrivial band
topologies of both $2$D and $3$D systems are characterized by the Chern
numbers and the $Z_{2}$ invariants, respectively, while the topological
property of bulk bands in $1$D periodic systems is characterized by the Zak
phase \cite{Zak}. However, the role of the Zak phase is subtle: The Zak
phase is not a geometric invariant, since it depends on the choice of origin
of the Brillouin zone. Only the Zak phase difference can identify a
topological transition. On the other hand, it has been shown that the
geometric phase can be complex \cite%
{Garrison,Dattoli,Ning,Moore,Pont,Massar,Ge,Whitney,Mostafazadeh} in a
non-Hermitian system. Motivated by the performance\ of the Zak phase in a
non-Hermitian system, in this work, we investigate the influence of
non-Hermitian term on the Zak phase in a bipartite lattice.

In this paper, we study the influence of non-Hermitian term on the Zak phase
of a Bloch system. We show exactly that for a bipartite system the real part
of the Zak phase cannot be affected by a staggered imaginary potential.
Comparing to a Hermitian system, a nonzero imaginary part appears in the Zak
phase, which amplifies/attenuates the Dirac norm of the evolved state. In
this sense, the Zak phase in a non-Hermitian system can still be used to
characterize the difference of two topological phases. We apply this result
to a non-Hermitian Su-Schrieffer-Heeger (SSH) model, in which the Zak phase
can be obtained by an adiabatic time evolution under the time-dependent
threading flux. It is shown that the difference of real part of Zak phases
for two different distortions are observable. To further demonstrate this
finding, we investigate the scattering problem for a time-dependent
scattering center, which is a magnetic-flux-driven non-Hermitian SSH ring.
Owing to the nature of the Zak phase, the intriguing features of this design
are the wave-vector independence and allow two distinct behaviors, perfect
transmission or confinement, depending on the timing of a flux impulse
threading the ring. When the flux is added during a wavepacket travelling
within the ring (scattering center), the wavepacket is confined in the
scatter partially. Otherwise, it exhibits perfect transmission through the
scatter. The result holds for wavepacket with arbitrary central momentum.

This paper is organized as follows. In Section \ref{sec_Zak}, we present a
general theory about the partial topological phase in $1$D non-Hermitian
bipartite system. In Section \ref{sec_SSH}, we apply the theory to a
concrete model and provide a dynamical method to realize Zak phase. Section %
\ref{sec_DY} devotes to the scattering behaviors based on the topological
feature of Zak phase. Finally, we give a summary and discussion in Section %
\ref{Summary}.

\section{Zak phase in a non-Hermitian bipartite system}

\label{sec_Zak} %==========================================================

\begin{figure}[tbp]
\centering
%\centering
%\includegraphics[ bb=160 500 407 680, width=0.35\textwidth, clip]{Fig_lb.eps}
\includegraphics[height=5cm, width=6.32cm]{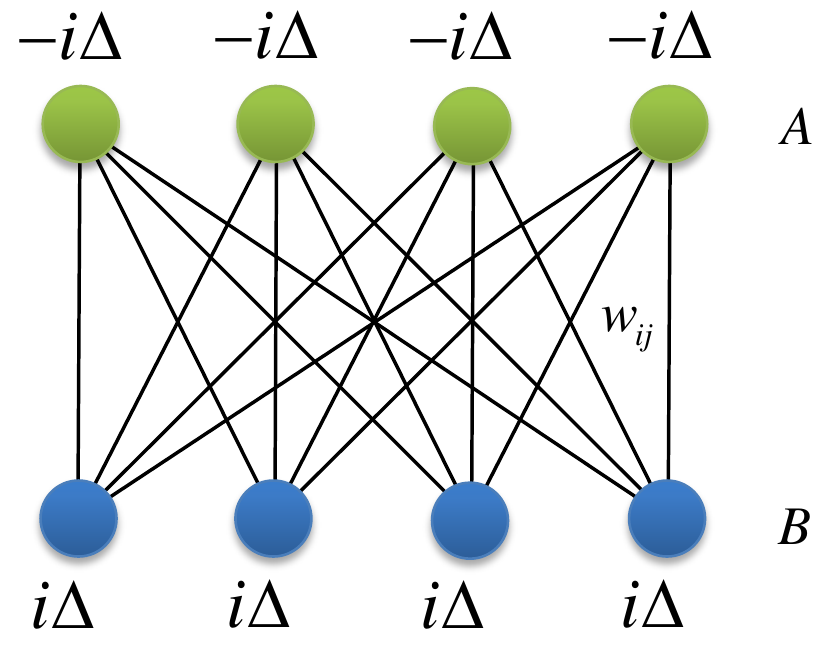}
%\centering
%\includegraphics[ bb=23 170 550 590, width=0.3\textwidth, clip]{fig1b.eps} %
%\includegraphics[ bb=23 170 550 590, width=0.3\textwidth, clip]{fig1cr.eps}
\caption{(Color online) Schematic illustration of the non-Hermitian
bipartite lattice that consists of two sublattices $A$ and $B$ with
identical lattice length. The two sublattices are connected with each other
by bond $w_{ij}$ which is across the $i$th site in sublattice $A$ and the $j$%
th site in sublattice $B$.}
\label{fig_lb}
\end{figure}

%================================================================

We first investigate the generic non-Hermitian lattice models that consists
of two sublattices, $A$ and $B$, the non-Hermiticity of which stems from the
staggered imaginary on-site potential. For clarity, we start discussion with
systems that possess the identical sublattice numbers $A=B=N$. The
corresponding bipartite non-Hermitian Hamiltonian can be written as
\begin{eqnarray}
\mathcal{H} &\mathcal{=}&\sum_{ij}w_{ij}c_{A,i}^{\dag }c_{B,j}+\text{\textrm{%
H.c.}}  \notag \\
&&-i\Delta \sum_{i}\left( c_{A,i}^{\dag }c_{A,i}-c_{B,i}^{\dag
}c_{B,i}\right) ,
\end{eqnarray}%
where $c_{A\left( B\right) ,j}^{\dagger }$ denotes the creation operator of
an electron on site $A$ ($B$) with periodic boundary condition $c_{A\left(
B\right) ,j}^{\dagger }=c_{A\left( B\right) ,j+N}^{\dagger }$ and $w_{ij}$
is a complex number describing the coupling constant between the two
sublattices. A schematic illustration of the model is presented in Fig. \ref%
{fig_lb}. Owing to the complexity of the coupling $w_{ij}$, the system does
not have the chirality-time-reversal symmetry but has translation symmetry
with the condition $w_{ij}=w_{i+1,j+1}$, i.e, $\left[ T,\mathcal{H}\right]
=0 $. Here the translation operator $T$ is defined as
\begin{equation}
T^{-1}c_{A\left( B\right) ,j}^{\dagger }T=c_{A\left( B\right) ,j+1}^{\dagger
},
\end{equation}%
which allows the invariant subspace spanned by the eigenvector of operator $%
T $. Taking the Fourier transformation, the non-Hermitian Bloch Hamiltonian
of a lattice with translational symmetry then reads $\mathcal{H=}%
\sum_{k}H_{k}$ satisfying $\left[ H_{k^{\prime }},H_{k}\right] =0$. In the
Nambu representation, the bipartite non-Hermitian Hamiltonian can be written
as
\begin{equation}
\mathcal{H=}\sum_{k}\mathcal{\eta }_{k}^{\dagger }h_{k}\mathcal{\eta }_{k},
\end{equation}%
where the basis $\mathcal{\eta }_{k}^{\dagger }=\left( c_{A,k}^{\dagger }%
\text{, }c_{B,k}^{\dagger }\right) $ with $c_{A\left( B\right) ,k}^{\dagger
} $ the creation operator of a Fermion in the momentum space, which
satisfies $T^{-1}c_{A\left( B\right) ,k}^{\dagger }T=e^{-ik}$, and%
\begin{equation}
h_{k}=\overrightarrow{B}\left( k\right) \cdot \overrightarrow{\sigma },
\label{hk}
\end{equation}%
with $\overrightarrow{\sigma }=\left( \sigma _{x},\sigma _{y},\sigma
_{z}\right) $ the vector of the matrices. Note that $\overrightarrow{B}%
\left( k\right) =\left( B_{x}\left( k\right) ,B_{y}\left( k\right)
,B_{z}\right) $ is a three-dimensional complex vector field, where $%
B_{z}=-i\Delta $ is $k$-independent. The presence of the staggered imaginary
on-site potential results in the imaginary strength of $z$ direction of $%
\overrightarrow{B}\left( k\right) $, i.e., $B_{z}^{\ast }=-B_{z}$. The
general energy expression of the single quasiparticle can be given as $%
\varepsilon _{k}=\pm r_{k}$, where $r_{k}=\sqrt{B_{x}^{2}\left( k\right)
+B_{y}^{2}\left( k\right) -\left[ \text{Im}\left( B_{z}\right) \right] ^{2}}$%
. It is clear that when any one of the qusimomenta $k$ satisfies $%
B_{x}^{2}\left( k\right) +B_{y}^{2}\left( k\right) -\left( \text{Im}\left(
B_{z}\right) \right) ^{2}<0$, the imaginary energy level appears in the
quasiparticle spectrum, which leads to the occurrence of complex energy
levels. This result has three implications. (i) The non-Hermitian
Hamiltonian $\mathcal{H}$ is pseudo-Hermitian, since its eigenvalues are
either real or come in complex-conjugate pairs. (ii) One can always modulate
the strength of imaginary on-site potential to obtain the full real
spectrum. The critical strength of the imaginary on-site potential depends
on the energy gap between the two bands of the Hermitian version with $%
B_{z}=0$. (iii) The EP occur at $B_{x}^{2}\left(
k\right) +B_{y}^{2}\left( k\right) =\left[ \text{Im}\left( B_{z}\right) %
\right] ^{2}$, which corresponds to the Jordan Block of $h_{k}$ accompanied
by the coalescence of the two eigenstates. Now we give the expression of the
eigenstates. The eigenstates of a bipartite non-Hermitian Hamiltonian can
construct a complete set of biorthogonal bases in association with the
eigenstates of its Hermitian conjugate. For the concerned bipartite system, $%
\left\vert \varrho _{+}^{k}\right\rangle $, $\left\vert \varrho
_{-}^{k}\right\rangle $ of $h_{k}$ and $\left\vert \chi
_{+}^{k}\right\rangle $, $\left\vert \chi _{-}^{k}\right\rangle $ of $%
h_{k}^{\dagger }$ are the biorthogonal bases of the single-quasiparticle
invariant subspace, which are explicitly expressed as%
\begin{eqnarray}
\left\vert \varrho _{+}^{k}\right\rangle &=&\left(
\begin{array}{c}
\cos \frac{\theta }{2}e^{-i\varphi } \\
\sin \frac{\theta }{2}%
\end{array}%
\right) ,\text{ }\left\vert \varrho _{-}^{k}\right\rangle =\left(
\begin{array}{c}
\sin \frac{\theta }{2} \\
-\cos \frac{\theta }{2}e^{i\varphi }%
\end{array}%
\right) ,  \label{Eigen_1} \\
\left\vert \chi _{+}^{k}\right\rangle &=&\left(
\begin{array}{c}
\cos \frac{\theta }{2}e^{i\varphi } \\
\sin \frac{\theta }{2}%
\end{array}%
\right) ^{\ast },\text{ }\left\vert \chi _{-}^{k}\right\rangle =\left(
\begin{array}{c}
\sin \frac{\theta }{2} \\
-\cos \frac{\theta }{2}e^{-i\varphi }%
\end{array}%
\right) ^{\ast }.  \label{Eigen_2}
\end{eqnarray}%
Here the vector field $\overrightarrow{B}\left( k\right) $ is represented in
terms of polar coordinates as%
\begin{equation}
\overrightarrow{B}\left( k\right) =r\left( \sin \theta \cos \varphi ,\sin
\theta \sin \varphi ,\cos \theta \right)
\end{equation}%
where
\begin{eqnarray}
r &=&\sqrt{B_{x}^{2}\left( k\right) +B_{y}^{2}\left( k\right) -\left( \text{%
Im}\left( B_{z}\right) \right) ^{2}}, \\
\cos \theta &=&\frac{B_{z}}{r},\text{ }\tan \varphi =\frac{B_{y}\left(
k\right) }{B_{x}\left( k\right) }.
\end{eqnarray}%
It is easy to check that biorthogonal bases $\left\{ \left\vert \varrho
_{\lambda }^{k}\right\rangle ,\left\vert \chi _{\lambda }^{k}\right\rangle
\right\} \left( \lambda =\pm \right) $ satisfy the biorthogonal and
completeness conditions,
\begin{equation}
\left\langle \varrho _{\lambda }^{k}\right\vert \left. \chi _{\lambda
^{\prime }}^{k^{\prime }}\right\rangle =\delta _{\lambda \lambda ^{\prime
}}\delta _{kk^{\prime }}\text{, }\sum_{\lambda k}\left\vert \varrho
_{\lambda }^{k}\right\rangle \left\langle \chi _{\lambda }^{k}\right\vert =I.
\label{BC_Condition}
\end{equation}%
Note that these properties are independent of the reality of the spectrum
and are generally satisfied except at the EP. In the absence of the
staggered imaginary on-site potential, we have $\left\vert \varrho _{\lambda
}^{k}\right\rangle =\left\vert \chi _{\lambda }^{k}\right\rangle $ with $%
\theta =\pi /2$ and the conditions (\ref{BC_Condition}) reduce to the Dirac
orthogonal and completeness conditions. In the following, we focus on the
system with full real spectrum. This is crucial to achieve the main
conclusion. To characterize the topological property of the energy band, we introduce
the modified Zak phase%
\begin{equation}
\mathcal{Z}_{\pm }\mathcal{=}\int_{0}^{2\pi }\mathcal{A}_{k,\pm }\mathrm{d}k,
\end{equation}%
where the Berry connection is given by
\begin{equation}
\mathcal{A}_{k,\pm }=i\left\langle \chi _{\pm }^{k}\right\vert \partial
_{k}\left\vert \varrho _{\pm }^{k}\right\rangle =\partial _{k}\varphi
\mathcal{A}_{\varphi ,\pm }+\partial _{k}\theta \mathcal{A}_{\theta ,\pm },
\end{equation}%
with
\begin{equation}
\mathcal{A}_{\varphi ,\pm }=i\left\langle \chi _{\pm }^{k}\right\vert
\partial _{\varphi }\left\vert \varrho _{\pm }^{k}\right\rangle ,\text{ }%
\mathcal{A}_{\theta ,\pm }=i\left\langle \chi _{\pm }^{k}\right\vert
\partial _{\theta }\left\vert \varrho _{\pm }^{k}\right\rangle .
\end{equation}%
The straightforward algebra shows that
\begin{eqnarray}
\mathcal{Z}_{\pm } &\mathcal{=}&\pm \frac{1}{2}\int_{\varphi \left( 0\right)
}^{\varphi \left( 2\pi \right) }\left( 1+\cos \theta \right) \mathrm{d}%
\varphi ,  \label{MZ} \\
&=&Z_{\pm }\pm i\frac{\text{Im}\left( B_{z}\right) }{2}\int_{\varphi \left(
0\right) }^{\varphi \left( 2\pi \right) }\frac{1}{r_{k}}\mathrm{d}\varphi ,
\end{eqnarray}%
where $Z$ denotes the Zak phase of the Hermitian system without staggered
imaginary potential, i.e., the Bloch Hamiltonian $h_{k}$ with $%
\overrightarrow{B}\left( k\right) =\left( B_{x}\left( k\right) ,B_{y}\left(
k\right) ,0\right) $. Comparing to the Hermitian version with $\theta =\pi
/2 $, the presence of the staggered imaginary potential does not alter the
real part of the Zak phase but brings about an extra imaginary part, which
amplifies the Dirac probability of the adiabatic evolved state. In this
sense, if Zak phase of the original Hermitian bipartite Hamiltonian is
topological then the modified Zak phase of the non-Hermitian version will
inherit this topological property through its real part. Such modified Zak
phase is therefore referred to as the partial topological Zak phase. In the
following section, we will demonstrate firstly the partial topological Zak
phase can be realized by a magnetic-flux-driven non-Hermitian SSH ring and
then apply it to a scattering problem.
%==========================================================

\begin{figure}[tbp]
\centering
%\hskip -3mm
%\includegraphics[ bb=70 460 450 730, width=0.45\textwidth, clip]{Fig1.eps} %\centering
\includegraphics[height=5cm,width=6.5218cm]{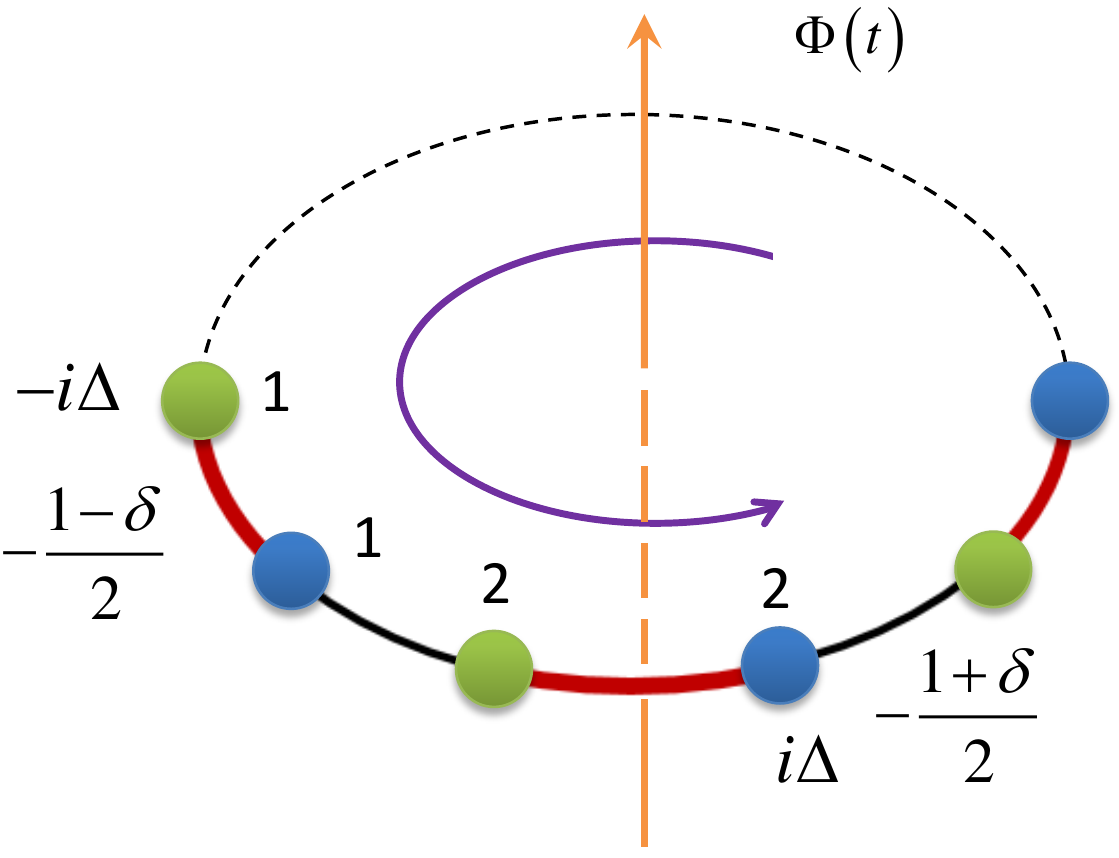}
\caption{(Color online) Schematic illustration of the non-Hermitian SSH
model driven by a time-dependent external field. The presence of the
magnetic field does not spoil the $\mathcal{PT}$ symmetry of the system. A
time-varying field $\Phi \left( t\right) $ induces the eddy field in a
direction indicated by the purple arrow, which acts as a linear field to
drive the wavepacket dynamics.}
\label{fig1}
\end{figure}

%================================================================
%==========================================================

\begin{figure*}[tbp]
\centering
%\centering
%\includegraphics[ bb=0 30 295 420, width=0.3\textwidth, clip]{Fig2a.eps} %
%\includegraphics[ bb=0 30 295 420, width=0.3\textwidth, clip]{fig2b.eps} %
%\includegraphics[ bb=0 30 295 420, width=0.3\textwidth, clip]{fig2c.eps}

\includegraphics[height=7.4cm, width=5.5cm]{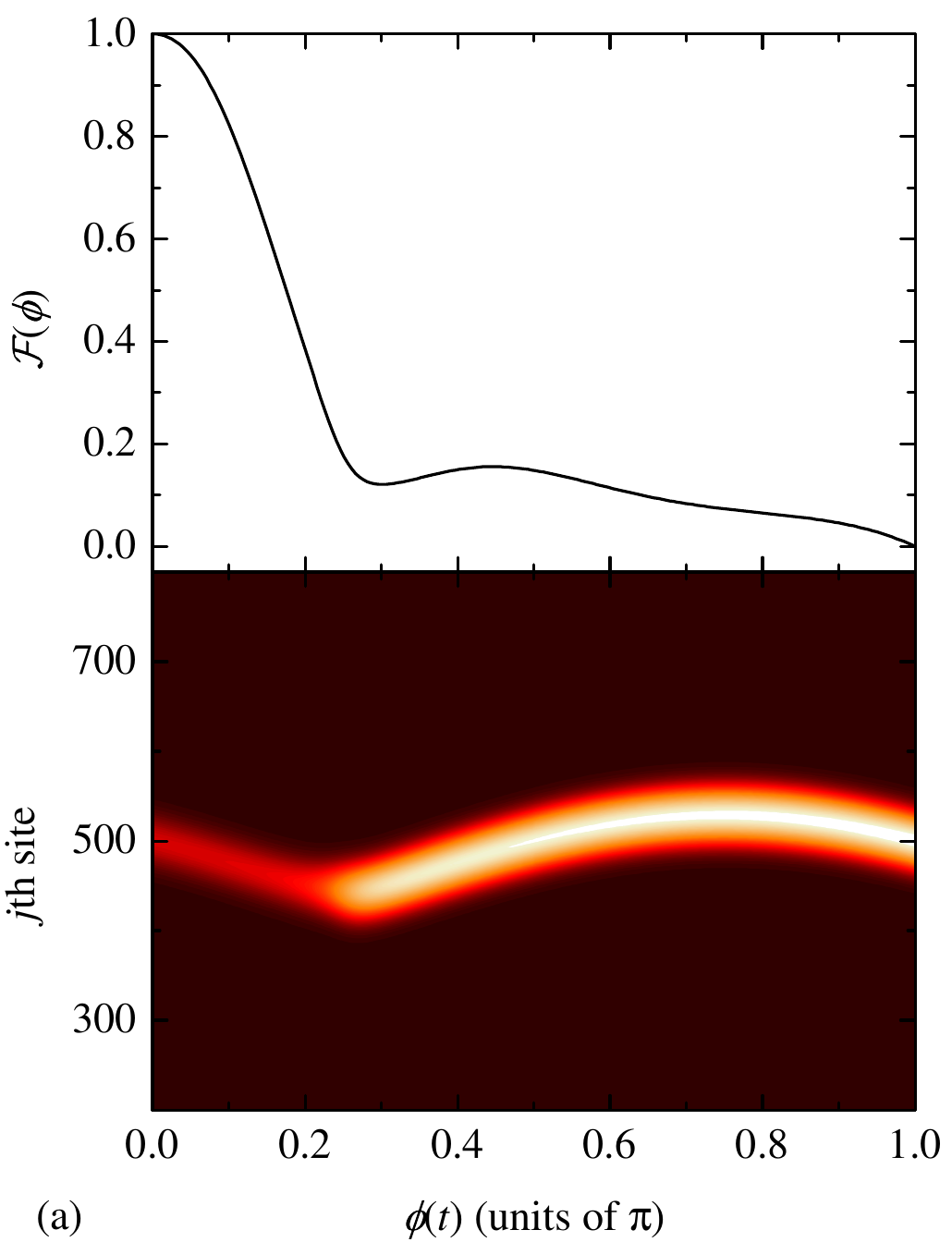}
\includegraphics[height=7.4cm, width=5.5cm]{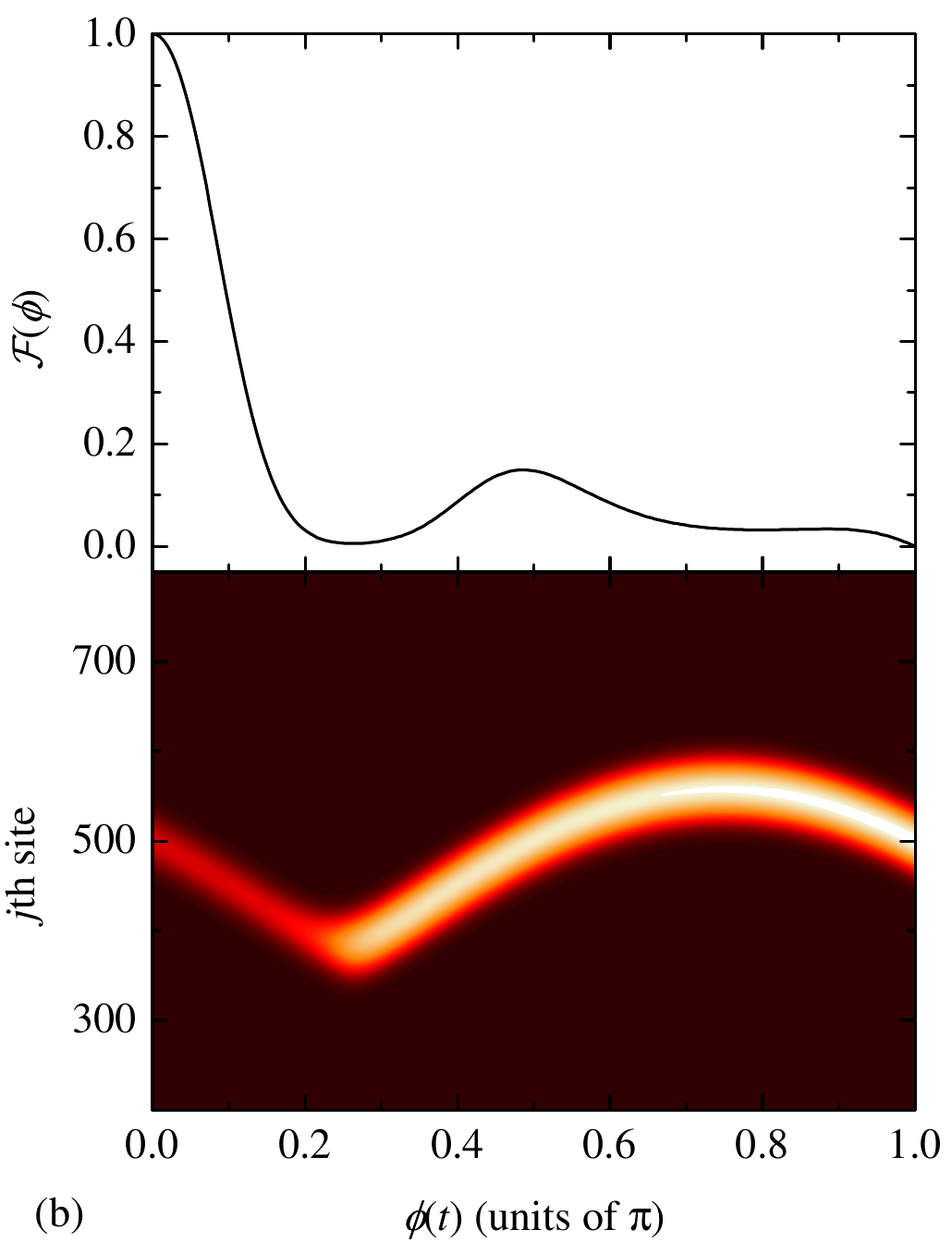}
\includegraphics[height=7.4cm, width=5.5cm]{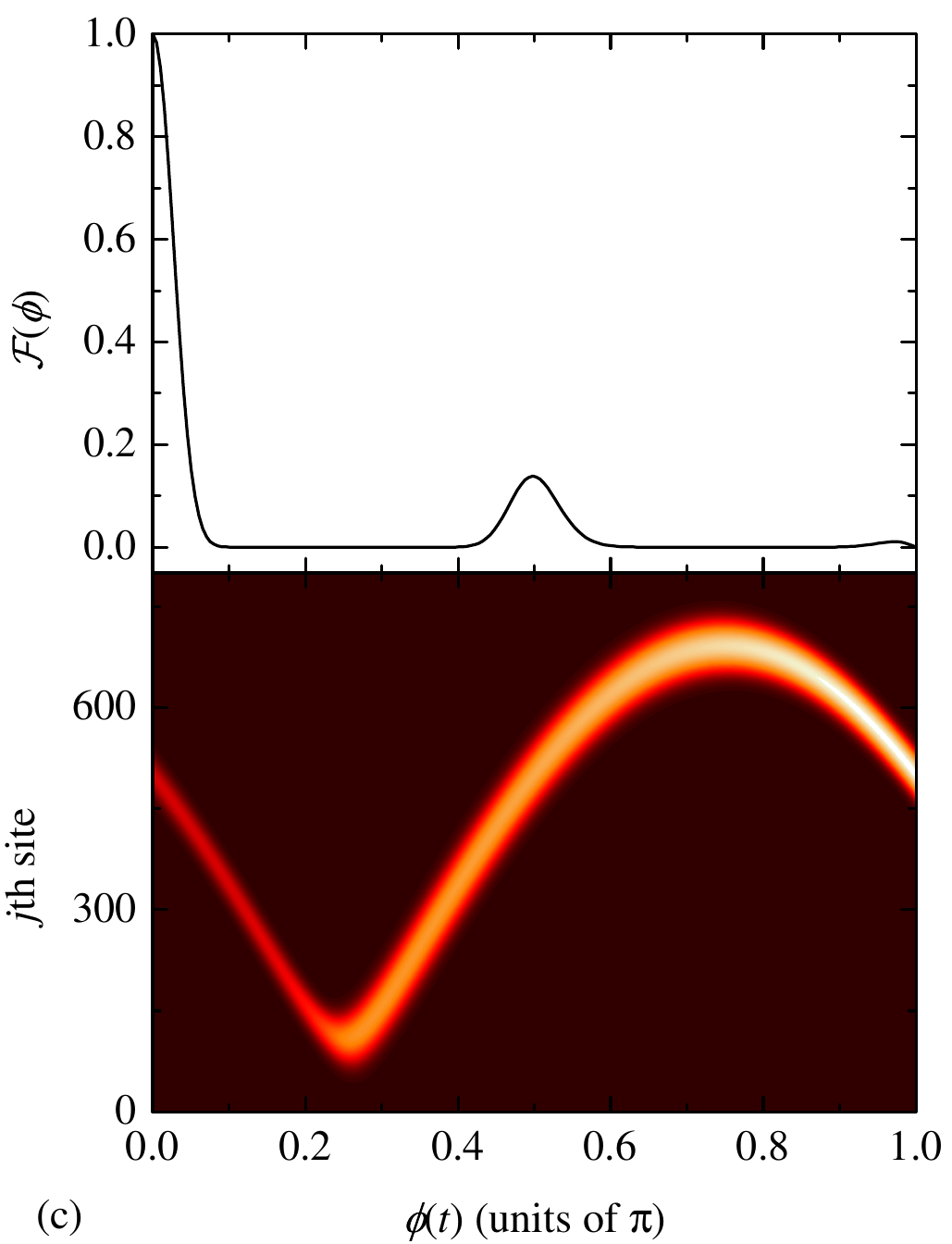}
\caption{(Color online) The profiles of the time evolution of a wave packet
in several typical cases. The initial wave packet is in the form of Eq. (%
\protect\ref{initial_WP}) with $g_{k}=\exp \left[ -\left( k-k_{0}\right)
^{2}/4\protect\alpha ^{2}-i\left( k-k_{c}\right) N_{c}\right] $, where the
half width of the wave packet $\protect\alpha =0.05$, central momentum $%
k_{0}=\protect\pi /4$ and the location of the initial state $N_{c}=250$. The
time is in units of $J^{-1}$, where $J$ is the scale of the Hamiltonian and
we take $J=1$. The other system parameters are $\protect\delta =0.15$, and $%
\Delta =0.1$. The magnetic flux adiabatically varies with (a) $\protect\beta %
=0.01$, (b) $\protect\beta =0.005$, and (c) $\protect\beta =0.0015$,
respectively. It can be shown that the wave packet exhibits half a BO with
different amplitude which is determined by $\protect\beta $. After half a
BO, the wave packet return back to the starting position. However, it is
orthogonal to the initial state that can be shown in the upper panel of each
subfigures. The numerical results demonstrate our analytical statement below
Eq. (\protect\ref{real_space}).}
\label{fig2}
\end{figure*}

%================================================================

\section{Non-Hermitian SSH model}

\label{sec_SSH} We consider a bipartite non-Hermitian SSH ring threaded by
magnetic flux, the Hamiltonian of which can be given as
\begin{eqnarray}
H &=&-\frac{1}{2}\sum_{j=1}^{2N}\left[ 1+\left( -1\right) ^{j}\delta \right]
\left( e^{i\phi }c_{j}^{\dag }c_{j+1}+\text{\textrm{H.c.}}\right)  \notag \\
&&+i\Delta \sum_{j}\left( -1\right) ^{j}c_{j}^{\dag }c_{j},  \label{H}
\end{eqnarray}%
the non-Hermiticity of which arises from the on-site staggered imaginary
potential $i\Delta \sum_{j}\left( -1\right) ^{j}c_{j}^{\dag }c_{j}$. The
system possesses a $2N$-site lattice, where $c_{j}$ is the annihilation
operator on site $j$ with the periodic boundary condition $c_{j+2N}=c_{j}$.
The nominal tunneling strength is staggered by $\delta $, and $\Phi =2N\phi $
is the magnetic flux threading the ring. We sketch the structure of the
system in Fig. \ref{fig1}. The origin Hermitian Hamiltonian with $\Delta =0$
can be realized with controlled defects using a system of attractive
ultracold fermions \cite{Chin,Strohmaier,Hacke} in a simple shaken
one-dimensional optical lattice. Furthermore, the non-Hermitian version can
be realized in a zigzag array of optical waveguides with alternating optical
gain and loss \cite{Longhi}. Before solving the Hamiltonian, it is
profitable to investigate the symmetry of the system and its breaking in the
eigenstates. Straightforward algebra shows that $\left[ \mathcal{PT},H\right]
=0$, that is, the Hamiltonian is $\mathcal{PT}$ symmetric even in the
presence of the magnetic flux, where the antilinear time-reversal operator $%
\mathcal{T}$ has the function $\mathcal{T}^{-1}i\mathcal{T}=-i$ and the
parity operator obeys $\mathcal{P}^{-1}c_{j}^{\dag }\mathcal{P}%
=c_{2N-j+1}^{\dag }$. However, the eigenstates does not always hold this
symmetry. According to the non-Hermitian quantum mechanics, the occurrence
of the EP always accomplishes the $\mathcal{PT}$ symmetry breaking of an
eigenstate. In the following, we will demonstrate this point.

We note that the Hamiltonian is invariant through a translational
transformation, i.e., $\left[ T,H\right] =0$, where $T$ is the shift
operator that defined as
\begin{equation}
T^{-1}c_{j}^{\dagger }T=c_{j+2}^{\dagger }\text{.}
\end{equation}%
This allows invariant subspace spanned by the eigenvector of operator $T$.
The single-particle eigenvector of $T_{2}$\ can be expressed as $%
c_{A,k}^{\dag }\left\vert 0\right\rangle $ and $c_{B,k}^{\dagger }\left\vert
0\right\rangle $, where%
\begin{eqnarray}
c_{A,k}^{\dag } &=&\frac{1}{\sqrt{N}}\sum_{j}e^{ik\left( j-1/2\right)
}c_{2j-1}^{\dag }, \\
c_{B,k}^{\dagger } &=&\frac{1}{\sqrt{N}}\sum_{j}e^{ikj}c_{2j}^{\dag },
\end{eqnarray}%
satisfying%
\begin{equation}
T^{-1}c_{A\left( B\right) ,k}^{\dag }T=e^{-ik}c_{A\left( B\right) ,k}^{\dag
}.
\end{equation}%
Here, $c_{A,k}^{\dag }$ and $c_{B,k}^{\dagger }$ are two kinds of creation
operators of bosons (or fermions), with $k=2\pi n/N$ ($n\in \lbrack 1,N]$),
representing the particles in odd and even sublattices. Then the Bloch
Hamiltonian $H_{k}$ can be given as $H_{k}=\mathcal{\eta }_{k}^{\dagger
}h_{k}\mathcal{\eta }_{k}$ where $h_{k}=\overrightarrow{B}\left( k\right)
\cdot \overrightarrow{\sigma }$ with the 3D vector field
\begin{eqnarray}
B_{x}\left( k\right) &=&-\cos \left( k/2+\phi \right) , \\
B_{y}\left( k\right) &=&-\delta \sin \left( k/2+\phi \right) , \\
B_{z} &=&-i\Delta .
\end{eqnarray}%
Accordingly, the eigenvalue of single quasiparticle can be obtain readily as
\begin{eqnarray}
\varepsilon _{k} &=&\pm r_{k},\text{ } \\
r_{k} &=&\sqrt{\cos ^{2}\left( k/2+\phi \right) +\delta ^{2}\sin ^{2}\left(
k/2+\phi \right) -\Delta ^{2}}.
\end{eqnarray}%
Here we want to point out that in the absence of $\Delta $, the energy gap
is $2\delta $, which determines the EP occurring at $\Delta
=\Delta _{c}=\delta $. The corresponding biorthogonal eigenstates can be
determined by Eqs. (\ref{Eigen_1})-(\ref{Eigen_2}), which can be expressed
as
\begin{eqnarray}
\left\vert \varrho _{+}^{k}\right\rangle &=&\cos \frac{\theta }{2}%
e^{-i\varphi }c_{A,k}^{\dag }\left\vert Vac\right\rangle +\sin \frac{\theta
}{2}c_{B,k}^{\dag }\left\vert Vac\right\rangle , \\
\left\vert \varrho _{-}^{k}\right\rangle &=&\sin \frac{\theta }{2}%
c_{A,k}^{\dag }\left\vert Vac\right\rangle -\cos \frac{\theta }{2}%
e^{i\varphi }c_{B,k}^{\dag }\left\vert Vac\right\rangle ,
\end{eqnarray}%
where $\left\vert Vac\right\rangle $ is the vacuum state of the fermion $%
c_{j}$, and$\ $%
\begin{equation}
\cos \theta =\frac{-i\Delta }{r_{k}},\text{ }\tan \varphi =\frac{\delta \sin
\left( k/2+\phi \right) }{\cos \left( k/2+\phi \right) }.
\end{equation}%
Applying the $\mathcal{PT}$ operator to the fermion operators and its vacuum
state $\left\vert Vac\right\rangle $ , we have
\begin{equation}
\left( \mathcal{PT}\right) ^{-1}c_{A\left( B\right) ,k}^{\dag }\mathcal{PT}%
=e^{-ik/2}c_{B\left( A\right) ,k}^{\dag }\text{, }
\end{equation}%
and
\begin{equation}
\mathcal{PT}\left\vert Vac\right\rangle =0,
\end{equation}%
which are available in both the broken and unbroken region. Owing to the
relation $\left[ \mathcal{PT},H\right] =0$, the eigenstate $\left\vert
\varrho _{\lambda }^{k}\right\rangle $ of $H$ for a real eigenvalue is
always the eigenstate of the symmetry operator $\mathcal{PT}$. However, the
coefficients $\cos \frac{\theta }{2}$ and $\sin \frac{\theta }{2}$
experience a transition as follows when the corresponding
single-quasiparticle energy $\varepsilon _{k}$ changes from real to
imaginary: We have $\left( \cos \frac{\theta }{2}\right) ^{\ast }=\sin \frac{%
\theta }{2}$ for real $\varepsilon _{k}$ and $\left( \cos \frac{\theta }{2}%
\right) ^{\ast }=\cos \frac{\theta }{2}$, and $\left( \sin \frac{\theta }{2}%
\right) ^{\ast }=\sin \frac{\theta }{2}$ for the imaginary $\varepsilon _{k}$%
. This leads to the conclusion that the eigenstate $\left\vert \varrho
_{\lambda }^{k}\right\rangle $ is not $\mathcal{PT}$ symmetric in the broken
region.

With the help of Eq. (\ref{MZ}), one can give directly the modified Zak
phase based on the analytical solution%
\begin{equation}
\mathcal{Z}_{\pm }=\pm \left( \frac{\pi \text{sgn}\left( \delta \right) }{2}%
-i\Delta \delta \int_{0}^{2\pi }\frac{1}{4r_{k}\left( r_{k}^{2}+\Delta
^{2}\right) }\mathrm{d}k\right) ,  \label{Zak_phase}
\end{equation}%
where sgn$\left( .\right) $ denotes the sign function. Note that we only
consider the case of the system with full real spectrum. There are two
features in the expression of $\mathcal{Z}_{\pm }$: (i) $\mathcal{Z}_{\pm }$
does not depend on the magnetic flux $\phi $ due to the relation $\mathcal{A}%
_{k,\lambda }=\mathcal{A}_{k+2\pi ,\lambda }$. (ii) The real part of $%
\mathcal{Z}_{\pm }$ is topological. Here we want to stress that the real
part of $\mathcal{Z}_{\pm }$ is not gauge invariance since the different
Fourier transformations can change its value. However, the difference of
real part between $\mathcal{Z}_{\pm }$ in the regions of $\delta >0$ and $%
\delta <0$ is gauge invariance. This property provide a way to adiabatically
control the scattering of the wavepacket dynamics in the following section.

Before starting the discussion of the wavepacket dynamics, we first connect
the magnetic-flux-driven Berry phase to the modified Zak phase $\mathcal{Z}%
_{\pm }$. To this end, we consider an adiabatic evolution, in which an
initial eigenstate evolves into the instantaneous eigenstate of the
time-dependent Hamiltonian. From Eq. (\ref{H}), we know that $H$ is a
periodic function of $\phi $, $H\left( \phi \right) $ $=H\left( \phi +2\pi
\right) $. Considering the time-dependent flux $\phi \left( t\right) $, any
eigenstate $\left\vert \varrho _{\lambda }^{k}\left( 0\right) \right\rangle $%
\ will return back to $\left\vert \varrho _{\lambda }^{k}\left( 0\right)
\right\rangle $\ if $\phi \left( t\right) $\ varies adiabatically from $0$
to $2\pi $, and the evolved state is the instantaneous eigenstate $%
\left\vert \varrho _{\lambda }^{k}\left( \phi \right) \right\rangle $. More
explicitly, the adiabatic evolution of the initial eigenstate $\left\vert
\varrho _{\lambda }^{k}\left( 0\right) \right\rangle $ under the
time-dependent Hamiltonian $H\left( \phi \left( t\right) \right) $ can be
expressed as%
\begin{eqnarray}
\left\vert \Psi _{\lambda }^{k}\left( \phi \right) \right\rangle &=&\mathcal{%
T}\exp \left[ -i\int_{0}^{t}H\left( t\right) \mathrm{d}t\right] \left\vert
\varrho _{\lambda }^{k}\left( 0\right) \right\rangle \\
&=&e^{i\left( \alpha _{k}^{\lambda }+\gamma _{k}^{\lambda }\right)
}\left\vert \varrho _{\lambda }^{k}\left( \phi \right) \right\rangle .
\notag
\end{eqnarray}%
Here $\alpha _{k}^{\lambda }\left( \phi \right) $ the dynamics phase and $%
\gamma _{k}^{\lambda }\left( \phi \right) $ the adiabatic phase have the
form
\begin{eqnarray}
\alpha _{k}^{\lambda }\left( \phi \right) &=&-\int\nolimits_{0}^{\phi
}\varepsilon _{\lambda }^{k}\left( \phi \right) \frac{\partial t}{\partial
\phi }\mathrm{d}\phi , \\
\gamma _{k}^{\lambda }\left( \phi \right) &=&\lambda \int\nolimits_{0}^{\phi
}\mathcal{A}_{\phi }\mathrm{d}\phi .
\end{eqnarray}%
where the Berry connection $\mathcal{A}_{\phi }=\delta /2r_{k}\left(
r_{k}+i\Delta \right) $ with $\mathcal{A}_{\phi }=\mathcal{A}_{\phi +\pi }$.
When the flux $\phi $ varies from $0$ to $\pi $, one can verify that the
adiabatic phase is $k$ independent, which is similar to the case in the
modified Zak phase. Correspondingly, the expression of adiabatic phase can
be given as
\begin{equation}
\gamma ^{\pm }=\pm \left( \frac{\pi \text{sgn}\left( \delta \right) }{2}%
-i\Delta \delta \int_{0}^{\pi }\frac{1}{2r_{k}\left( r_{k}^{2}+\Delta
^{2}\right) }\mathrm{d}\phi \right) .
\end{equation}%
For a Hermitian system, the adiabatic phase is always real that ensures the
probability preserving evolution, while the probability of an evolved state
changes due to the imaginary part of the adiabatic phase in a non-Hermitian
system. The attenuation or amplification of probability depends on the sign
of the imaginary phase. Straightforward algebra shows that the imaginary
part of $\mathcal{Z}_{\pm }$ is the same as the imaginary part of $\gamma
^{\pm }$. This is always true for any values of $k$ $\left( \phi \right) $
for $\mathcal{Z}_{\pm }$ $\left( \gamma ^{\pm }\right) $. In this sense, one
can mimic the modified Zak phase $\mathcal{Z}_{\pm }$ through the adiabatic
variation of magnetic flux from $0$ to $\pi $. Here we want to point out
that although the magnetic-flux-driven adiabatic phase is identical to the
modified Zak phase, the evolve state $\left\vert \Psi _{\lambda }^{k}\left(
\pi \right) \right\rangle $ does not return back to the initial state. One
can readily obtain the $\left\vert \Psi _{\lambda }^{k}\left( \pi \right)
\right\rangle $ by modulating $\varphi \rightarrow \varphi +\pi $ in initial
state $\left\vert \varrho _{\lambda }^{k}\left( 0\right) \right\rangle $. In
the coordinate space, it can be achieved through modulating $\pi $ phase of
the distribution on the odd site, that is%
\begin{eqnarray}
\left\vert \Psi _{\lambda }^{k}\left( \pi \right) \right\rangle
&=&e^{i\alpha _{k}^{\lambda }\left( \pi \right) }e^{i\gamma ^{\lambda
}}\left\vert \varrho _{\lambda }^{k}\left( \pi \right) \right\rangle ,
\label{pi_relation} \\
&=&\frac{e^{i\alpha _{k}^{\lambda }\left( \pi \right) }e^{i\gamma ^{\lambda
}}}{\sqrt{N}}\sum_{j}\left( -\cos \frac{\theta }{2}e^{-i\varphi }e^{ik\left(
j-1/2\right) }c_{2j-1}^{\dag }\right.  \notag \\
&&\left. +\sin \frac{\theta }{2}e^{ikj}c_{2j}^{\dag }\left\vert
Vac\right\rangle \right) .  \notag
\end{eqnarray}%
This is crucial step to understand the wavepacket dynamics in the following.
%==========================================================

\begin{figure}[tbp]
\centering
%\centering
%\includegraphics[ bb=0 32 293 418, width=0.45\textwidth, clip]{Fig3.eps}
\includegraphics[height=10.46cm,width=8cm]{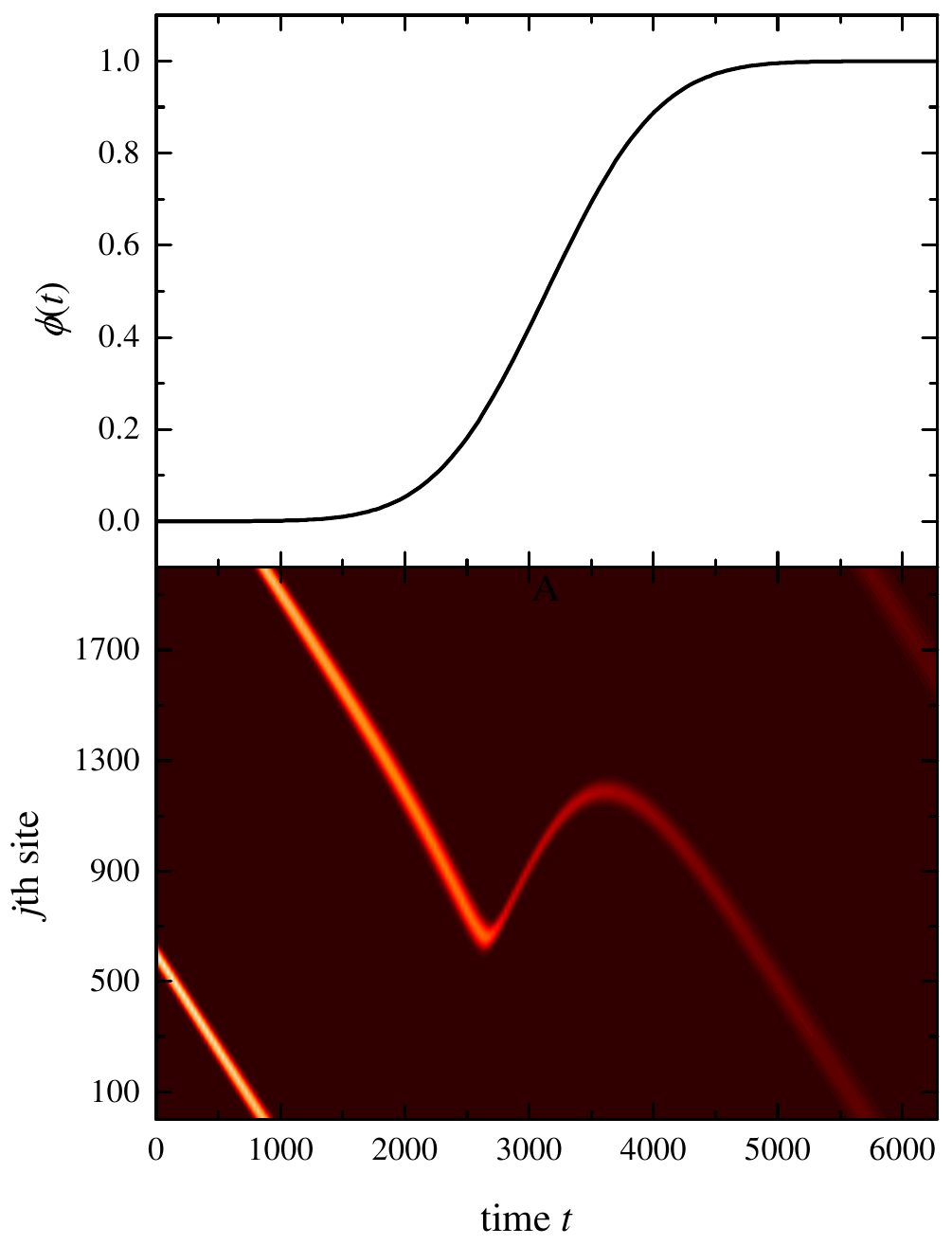}
\caption{(Color online) Numerical simulation of the evolved wavepacket
driven by the magnetic flux $\protect\phi \left( t\right) $ that varies as
the error function. The initial state and the system parameters are the same
with that of the Fig. \protect\ref{fig2}(c) except that $\protect\delta $ is
replaced with $-\protect\delta $. Hence the probability of the evolved
wavepacket is attenuated as time increases.}
\label{fig3}
\end{figure}

%================================================================
Now switch gear to the adiabatic time evolution of the wavepacket. We
consider two kinds of functions $\phi \left( t\right) $. For the first one,
the magnetic flux $\phi $ linearly depends on time, that is $\phi =\beta t$.
When the flux $\phi $ varies from $0$ to $\pi $, the dynamics phase $\alpha
_{k}^{\lambda }\left( \pi \right) $ is $k$ independent, which can be
verified through the fact that $\varepsilon _{k}\left( \phi \right) =$ $%
\varepsilon _{k}\left( \phi +\pi \right) $. Therefore, if one consider the
wavepacket dynamics, the adiabatic phase and dynamics phase are served as an
overall phase and cannot induce the interference among the instantaneous
eigenstates. More explicitly, we consider the wavepacket localized on the
upper band of the system (the conclusion is also hold for the case of lower
band)%
\begin{equation}
\left\vert G_{k_{0}}^{N_{A}}\left( 0\right) \right\rangle
=\sum_{k}g_{k}\left\vert \varrho _{+}^{k}\left( 0\right) \right\rangle .
\label{initial_WP}
\end{equation}%
where $N_{A}$ and $k_{0}$ denote the center and velocity of the initial
wavepacket, respectively. Here, we do not give the explicit expression of
the coefficient $g_{k}$, since the following analysis is irrelevant to $%
g_{k} $. In the coordinate space, the wavepacket can be expressed as
\begin{equation}
\left\vert G\left( 0\right) \right\rangle =\sum_{j}f_{j}c_{j}^{\dag
}\left\vert Vac\right\rangle ,
\end{equation}%
where the scripts $N_{A}$ and $k_{0}$ are neglected. Through an adiabatic
evolution in which $\phi $ varies from $0$ to $\pi $, the adiabatic phase
and dynamics phase is an overall phase then we have
\begin{equation}
\left\vert G\left( \pi \right) \right\rangle =e^{i\Omega _{+}}e^{\xi
_{+}}\sum_{k}g_{k}\left\vert \varrho _{+}^{k}\left( \pi \right)
\right\rangle .
\end{equation}%
where $\Omega _{+}=\alpha ^{+}\left( \pi \right) +$ Re$\left( \gamma
^{+}\right) $ and $\xi _{+}=$ $-$Im$\left( \gamma ^{+}\right) $. Owing to
the relation (\ref{pi_relation}), the evolved wavepacket at time $t=\pi
/\beta $ in the coordinate space can be given as
\begin{equation}
\left\vert G\left( \pi \right) \right\rangle =e^{i\Omega _{+}}e^{\xi
_{+}}\sum_{j}\left( -1\right) ^{j}f_{j}c_{j}^{\dag }\left\vert
Vac\right\rangle .  \label{adiabatic_evolve}
\end{equation}%
where the odd site acquires a phase $\pi $. It indicates that the two
wavepackets $\left\vert G\left( 0\right) \right\rangle $ and $\left\vert
G\left( \pi \right) \right\rangle $ are orthogonal based on
\begin{equation}
\mathcal{F}\left( \pi \right) =\frac{\left\langle G\left( 0\right)
\right\vert \left. G\left( \pi \right) \right\rangle }{\sqrt{\left\langle
G\left( 0\right) \right\vert \left. G\left( 0\right) \right\rangle
\left\langle G\left( \pi \right) \right\vert \left. G\left( \pi \right)
\right\rangle }}=0.
\end{equation}%
To demonstrate this feature, we plot the trajectories of the wave packet
with different $\beta $ in Fig. \ref{fig2}(a)-(c). It can be shown that the
wavepacket experiences half a Bloch oscillation (BO) accompanied by the
probability amplification in the coordinate space. The dynamics of a wave
packet driven by time-dependent magnetic flux is the same as that driven by
a linear field with strength $\beta $, according to the quantum Faraday's
law \cite{HWH}. Furthermore, the center path of a wave packet driven by a
linear field accords with the dispersion of the Hamiltonian in the absence
of the field within the adiabatic regime \cite{LS}%
\begin{equation}
x_{c}\left( \phi \right) =x_{c}\left( 0\right) +\frac{1}{\beta }\left[
\varepsilon _{k_{c}}\left( \phi \right) -\varepsilon _{k_{c}}\left( 0\right) %
\right] ,  \label{real_space}
\end{equation}%
where $\varepsilon _{k_{c}}\left( \phi \right) $ is the dispersion relation
and $k_{c}$ is the central momentum of the wave packet. From this
perspective, the amplitude of the BO of wavepacket is inversely proportional
to $\beta $.

For the second one, the magnetic flux varies with time according to the
error function curve, that is $\phi =$ erf$\left( t\right) $, where erf$%
\left( .\right) $ is the error function. In this situation, the dynamics
phase $\alpha _{k}^{\lambda }\left( \pi \right) $ is $k$ dependent but
adiabatic phase $\gamma ^{\pm }$ is still $k$ independent. Therefore the
dynamics of the wavepacket is more complicated than the case of $\phi =\beta
t$. In Fig. \ref{fig3}, we plot the trajectory of the wavepacket. On the one
hand, we can see that when $\mathrm{d}\phi /\mathrm{dt}$ is small which
corresponds to two ends of the error function, the wavepacket travels at
approximately uniform speed. In this condition, the effective linear field
with strength $\beta $ is approximate zero. On the other hand, the
derivative $\mathrm{d}\phi /\mathrm{dt}$ is linear in the middle of the
error function. Therefore, there exists a linear field that drives the
wavepacket oscillate in the coordinate space, which can be shown in Fig. \ref%
{fig3}. For the sake of simplicity, we consider the first case that $\phi
=\beta t$ to realize the dynamical control of the wavepacket.

\section{Transmission and confinement}

\label{sec_DY} In this section, we will control the scattering behavior of
the wavepacket based on the partial topological property of the Zak phase.
We will show that the wavepacket will display two distinct dynamical
behaviors in the modulating non-Hermitian scattering network, that is
perfect transmission and partial confinement.

\subsection{Interferometer}

\label{sec_Interferometer}
%==========================================================

\begin{figure}[tbp]
\centering
%\centering
%\includegraphics[ bb=19 416 567 623, width=0.48\textwidth,
%clip]{Fig_setup.eps}
\includegraphics[height=3cm,width=8.45cm]{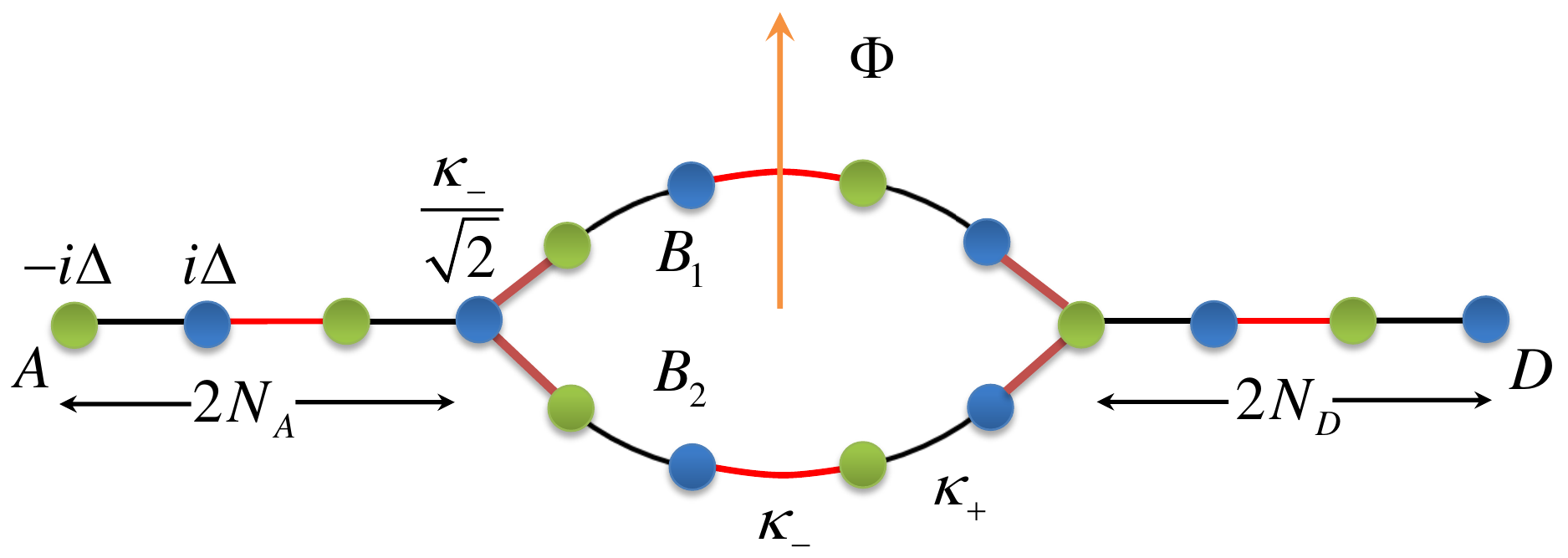}
\caption{(Color online) Schematic illustration of the scattering setup. The
scattering system consists of three components: an input non-Hermitian SSH
lead $A$ with length $N_{A}$, an output non-Hermitian SSH lead $D$ with
length $N_{D}$, and a ring $\{B_{1}$, $B_{2}\}$ threaded by a magnetic flux $%
\Phi \left( t\right) =4\left( N_{B}+1/2\right) \protect\phi \left( t\right) $%
, where $N_{B}$ represents the length of $B_{1\left( 2\right) }$. Here $%
\protect\kappa _{+}=-\left( 1-\protect\delta \right) /2$, and $\protect\kappa %
_{-}=-\left( 1+\protect\delta \right) /2$. The sub chains $B_{1}$ and $B_{2}$
are $\mathcal{P}$ symmetric with respect to the direction of incident GWP.
Note that the hopping constants connecting the leads and the scattering ring
are modulated with $\protect\kappa _{-}/\protect\sqrt{2}$, which ensure that
the network can be decoupled into two independent virtual chains with
different length. In the absence of the magnetic flux, therefore, the whole
propagation process in the real space is as follows: When the initial GWP
reaches the node, it is divided into the two identical GWPs which also move
with same speed along the legs $B_{1}$ and $B_{2}$ respectively without
spreading. In the scattering center, the upper and lower GWPs are driven by
the effective Hamiltonians which can be constructed by extending the two
legs to the completed non-Hermitian SSH rings. The difference between two
effective Hamiltonians is the sign of $\protect\delta $ resulting from two
symmetric legs. This also indicates that when the two GWPs experience half a
BO, they acquire the phase difference of $\protect\pi $.}
\label{fig_setup}
\end{figure}

%================================================================
In order to demonstrate these behaviors, we first consider the
interferometer model which is illustrated schematically in Fig. \ref%
{fig_setup}. This quantum interferometer consists of two non-Hermitian SSH
chains $A$, $D$ and a ring $\left\{ B_{1},B_{2}\right\} $ threaded by
magnetic flux in the unit of flux quanta. The corresponding Hamiltonian reads%
\begin{equation}
H_{\mathrm{net}}=-\frac{1}{2}\sum_{\alpha }\left( H_{\alpha }+H_{\mathrm{%
joint}}\right) ,  \label{H_S}
\end{equation}%
where $\alpha =A$, $B_{1}$, $B_{2}$, $D$ denote four non-Hermitian SSH
chains, respectively, and
\begin{eqnarray}
H_{\sigma _{1}} &=&\sum_{j=1}^{2N_{\sigma _{1}}-1}\left[ 1+\left( -1\right)
^{j}\delta \right] \left( c_{\sigma _{1},j}^{\dag }c_{\sigma _{1},j+1}+\text{%
\textrm{H.c.}}\right)  \notag \\
&&-2i\Delta \sum_{j=1}^{2N_{\sigma _{1}}-1}\left( -1\right) ^{j}c_{\sigma
_{1},j}^{\dag }c_{\sigma _{1},j}, \\
H_{\sigma _{2}} &=&\sum_{j=1}^{2N_{\sigma _{2}}-1}\left[ 1+\left( -1\right)
^{j}\delta \right] \left( e^{i\phi }c_{\sigma _{2},j}^{\dag }c_{\sigma
_{2},j+1}+\text{\textrm{H.c.}}\right)  \notag \\
&&-2i\Delta \sum_{j=1}^{2N_{\sigma _{2}}-1}\left( -1\right) ^{j}c_{\sigma
_{1},j}^{\dag }c_{\sigma _{1},j},
\end{eqnarray}%
where $\sigma _{1}=A,$ $D$ and $\sigma _{2}=B_{1}$, $B_{2}$. Note that $%
H_{B_{1}}$ and $H_{B_{2}}$ describe the two identical non-Hermitian SSH
chains with length $N_{B}\equiv N_{B_{1}}=N_{B_{2}}$. The connection
Hamiltonian reads%
\begin{eqnarray}
H_{\mathrm{joint}} &=&\frac{\left( 1+\delta \right) }{\sqrt{2}}\left(
e^{i\phi }c_{A,2N_{A},}^{\dag }c_{B_{1},1}\right.  \notag \\
&&+e^{-i\phi }c_{a,2N_{A},}^{\dag }c_{B_{2},1}+e^{i\phi
}c_{B_{1},2N_{B},}^{\dag }c_{D,1}  \notag \\
&&\left. +e^{-i\phi }c_{B_{2},2N_{B},}^{\dag }c_{D,1}+\text{\textrm{H.c.}}%
\right) .
\end{eqnarray}%
Here $\Phi =4\left( N_{B}+1/2\right) \phi $ is the total magnetic flux
threading the ring.\ Now we focus on the dynamics of the wavepacket based on
the partial topological property of the modified Zak phase. To this end, we
take the initial state as the Gaussian wavepacket (GWP)
\begin{equation}
\left\vert G\left( k_{0},0\right) \right\rangle =\frac{1}{\sqrt{\Omega }}%
\sum_{l=1}^{2N_{A}}e^{-\alpha ^{2}\left( l-N_{c}\right)
^{2}}e^{ik_{0}l}c_{A,l}^{\dag }\left\vert Vac\right\rangle ,
\end{equation}%
with the central momentum $k_{0}$. Here, $\Omega $ is the normalization
factor and $N_{c}\in \left[ 1,2N_{A}\right] $ is the initial central
position of the GWP at the input chain $A$ while the factor $\alpha $ is
large enough to guarantee the locality of the state in the chain $A$.
According to the Ref. \cite{HWHPRA}, when the center momentum $k_{0}$
satisfies the condition $\left\vert k_{0}+\pi /2\right\vert \gg 0$ and $%
\delta $ is a small number, the initial state will distribute on $k\sim
2k_{0}$ in the upper band of the Hamiltonian $H_{A}$ with periodic boundary
condition. It is worthy pointing out that when $k_{0}=\pi /4,$ $3\pi /8$ and
$\pi /2$, such a GWP can approximately propagate along the non-Hermitian SSH
chain without spreading \cite{HWHPRA}. For simplicity, the center momentum $%
k_{0}$ is assumed to be $\pi /4$ in the following, which ensures that the
initial state is mainly localized on the upper band of the Hamiltonian $%
H_{A} $ with either $\delta >0$ or $\delta <0$.
%==========================================================

\begin{figure*}[tbp]
\centering
%\centering
%\includegraphics[ bb=0 22 317 417, width=0.32\textwidth, clip]{Fig5a.eps} %
%\includegraphics[ bb=0 22 317 417, width=0.32\textwidth, clip]{Fig5b.eps} %
%\includegraphics[ bb=40 290 270 745, width=0.2\textwidth, clip]{Fig5c.eps}
\includegraphics[height=7.53cm,width=6cm]{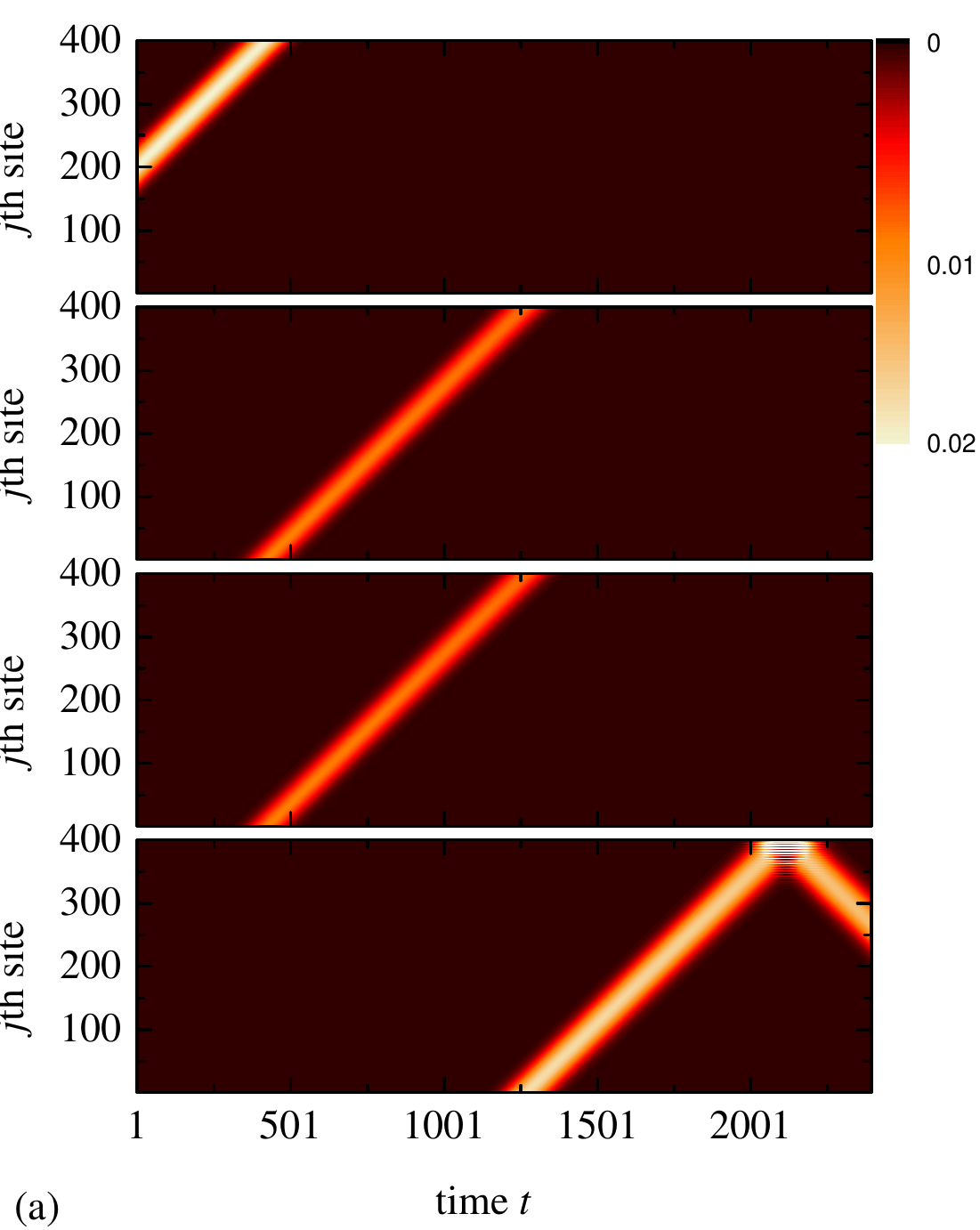}
\includegraphics[height=7.53cm,width=6cm]{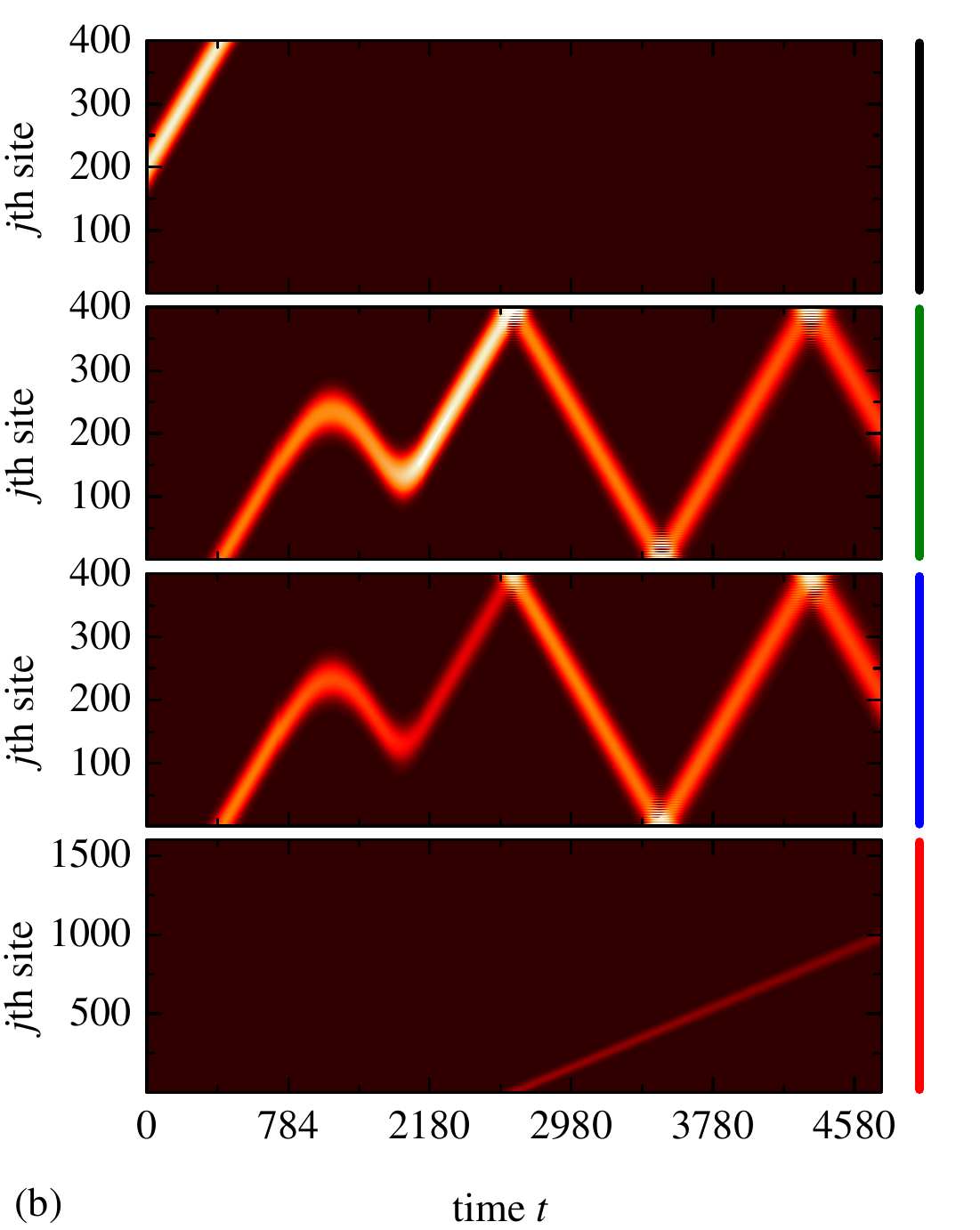}
\includegraphics[height=7.53cm,width=4.02cm]{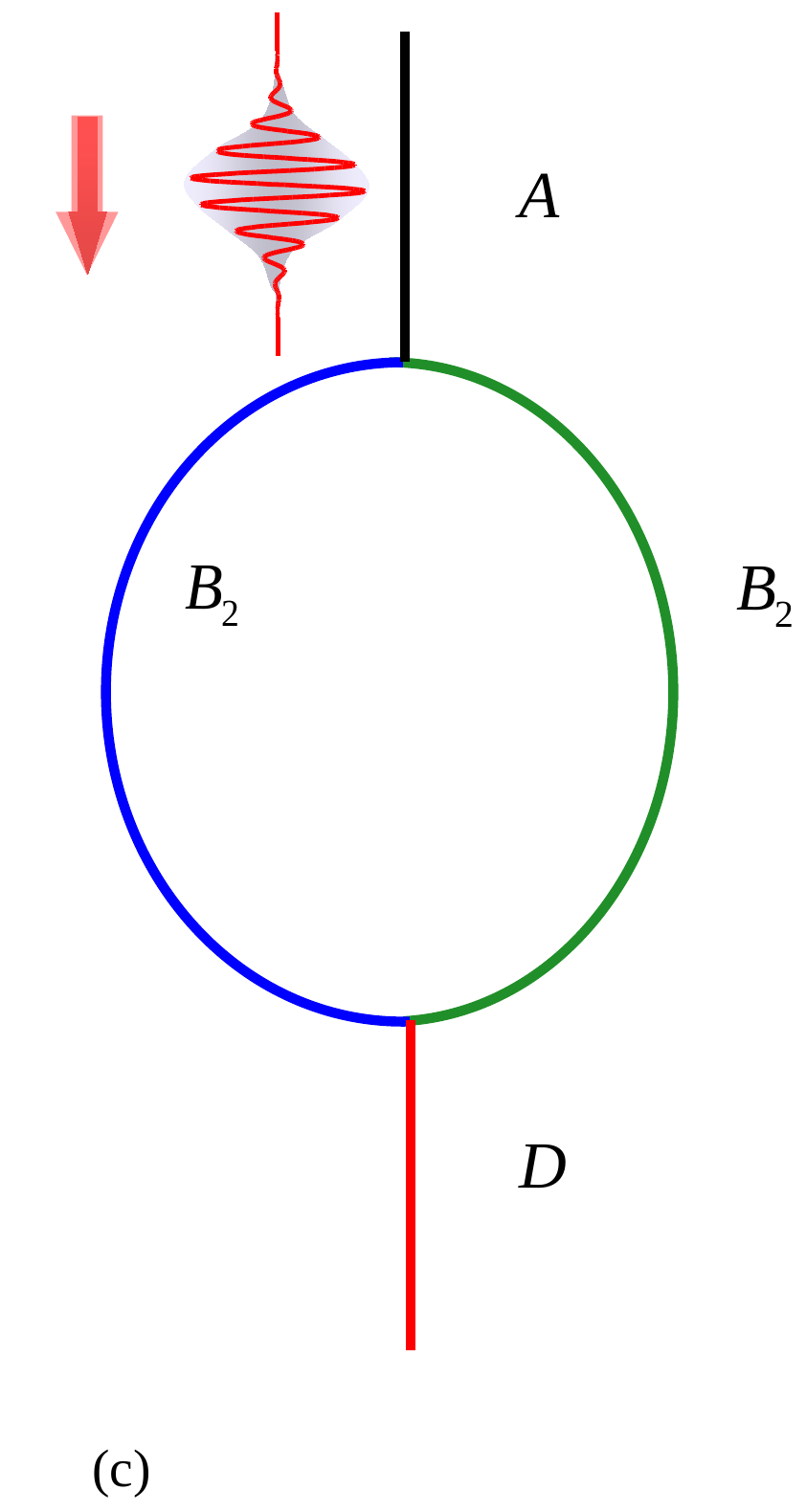}
\caption{(Color online) Propagation of the GWP in the non-Hermitian
scattering system with (a) $N_{A}=N_{B}=N_{D}=400$, and (b) $%
N_{A}=N_{B}=400$, $N_{D}=1600$. The other system parameters are $\Delta =0.5$, and $\delta =0.15$. (a) describes the case that the magnetic flux has changed from $0$ to $\protect\pi $ before the GWP enters
into the scattering ring. In this condition, the GWP travels along the
virtual chain $a$ as shown in the appendix. Therefore the GWP can pass
perfectly through the scattering center with no reflection at the two nodes.
In this sense, the non-Hermitian scattering ring is invisible to the
incident GWP. (b) When the GWP enters into the scattering ring, the magnetic
flux is switched on. In this case, each of the two cloned GWP undergoes half
a BO. The probability of upper GWP is amplified due to $\protect\delta >0$
of the effective Hamiltonian. On the contrary, the probability of lower GWP
is attenuated driven by the corresponding effective Hamiltonian with $%
\protect\delta <0$. The adiabatic process bring about a phase difference $%
\protect\pi $ between two such cloned GWPs. Therefore, the partial
probability of two cloned GWPs is confined in the scattering center, which
is inaccordance with our theoretical prediction. (c) Schematic illustration
of the concerned scattering system. The four panels of subfigures (a) and
(b) represent the input lead $A$, scattering center $\left\{ B_{1}\text{, }%
B_{2}\right\} $ and output chain $D$, which are denoted by black, green,
blue, and red lines, respectively. Note that the scale of the fourth panel
of subfigure (b) is different from the other panels.}
\label{fig5}
\end{figure*}
%================================================================
Owing to the Eq. (\ref{t_1}) of the appendix, the initial GWP will travel
along the virtual chain $a$ in the absence of the magnetic flux. Actually,
at a certain time $\tau $, such GWP evolves approximately into%
\begin{equation}
\left\vert G\left( \frac{\pi }{4},\tau \right) \right\rangle \sim \frac{1}{%
\sqrt{\Omega }}\sum_{l=2N_{A}+1}^{2N_{A}+2N_{B}}e^{-\alpha ^{2}\left(
l-N_{c}-2v\tau \right) ^{2}}e^{i\frac{\pi }{4}l}\widetilde{c}_{a,l}^{\dag
}\left\vert Vac\right\rangle
\end{equation}%
in the virtual space, where $v=\left\vert \left( \partial \varepsilon
_{k}/\partial k\right) _{\pi /2}\right\vert $ represents the group velocity
of the GWP. From the mapping of the operators (\ref{t_1})-(\ref{t_4}), we
have the final state as%
\begin{equation}
\left\vert G\left( \frac{\pi }{4},\tau \right) \right\rangle =\frac{1}{\sqrt{%
2}}\sum_{p=1}^{2}\left\vert G_{p}\left( \frac{\pi }{4},\tau \right)
\right\rangle ,
\end{equation}%
where%
\begin{equation}
\left\vert G_{p}\left( \frac{\pi }{4},\tau \right) \right\rangle =\frac{1}{%
\sqrt{\Omega }}\sum_{j=1}^{2N_{B}}e^{-\alpha ^{2}\left( j-N_{\tau }\right)
^{2}}e^{i\frac{\pi }{4}j}c_{B_{p},j}^{\dag }\left\vert Vac\right\rangle
\end{equation}%
is the clone of the initial GWP with the center $N_{\tau }\in \left[ 1,2N_{B}%
\right] $. The beam splitter split the single-particle GWP into $2$ cloned
GWPs without any reflection. Now we investigate the effect of the magnetic
flux threading the ring adiabatically on the dynamics of GWP. We consider
the two cases:

\subsection{Perfect transmission}

\label{sec_PT} We first consider the case that the magnetic flux $\phi $ has
changed from $0$ to $\pi $ before the GWP enters into the ring. Under this
condition, the GWP cannot feel the presence of magnetic flux. Therefore it
will travel along the virtual chain $a$ without any reflection. The final
state in the virtual space can be expressed as
\begin{equation}
\left\vert G_{p}\left( \frac{\pi }{4},\tau _{f}\right) \right\rangle =\frac{1%
}{\sqrt{\Omega }}\sum_{l=2N_{A}+2N_{B}+1}^{N}e^{-\alpha ^{2}\left(
l-N_{f}\right) ^{2}}e^{i\frac{\pi }{4}l}\widetilde{c}_{a,l}^{\dag
}\left\vert Vac\right\rangle ,
\end{equation}%
with $N_{f}\in \left[ 2N_{A}+2N_{B},\text{ }N\right] $. We detail this
process in the appendix section. In the coordinate space, the GWP will pass
perfectly through the scattering center. We plot the Fig. \ref{fig5}(a) to
demonstrate this case.

\subsection{Partial confinement}

\label{sec_PC} Second, we consider the case that the magnetic flux $\phi $
is varied from $0$ to $\pi $ during a wavepacket travelling within the ring.
In this situation, the initial GWP $\left\vert G\left( \frac{\pi }{4}%
,0\right) \right\rangle $ first enters into the ring so that it is split
into two cloned GWPs at time $\tau $. When the magnetic flux is switched on,
the two cloned GWPs will experience half a BO. However, the corresponding
effective driven Hamiltonians are different for two cloned GWPs. For the
upper GWP, the effective driven Hamiltonian can be obtained through
extending the Hamiltonian $H_{B_{1}\text{ }}$ to a complete SSH ring. On the
other hand, for a lower GWP, one can check that the effective driven
Hamiltonian can be constructed by replacing $\delta $ of the upper effective
SSH ring with $-\delta $. Therefore, the two cloned GWPs acquire two
opposite adiabatic phase after half a BO. The evolved state can be obtained
with the help of Eq. (\ref{adiabatic_evolve}) as
\begin{eqnarray}
\left\vert G\left( \frac{\pi }{4},\tau +\tau _{\mathrm{BO}}\right)
\right\rangle &=&e^{\xi _{+}}\left\vert \widetilde{G}_{1}\left( \frac{\pi }{4%
},\tau +\tau _{\mathrm{BO}}\right) \right\rangle  \notag \\
&&-e^{-\xi _{+}}\left\vert \widetilde{G}_{2}\left( \frac{\pi }{4},\tau +\tau
_{\mathrm{BO}}\right) \right\rangle  \label{ABO_GW}
\end{eqnarray}%
where $\tau _{\mathrm{BO}}$ represents the time that the GWP undergoes half
a BO and
\begin{eqnarray}
\left\vert \widetilde{G}_{p}\left( \frac{\pi }{4},\tau +\tau _{\mathrm{BO}%
}\right) \right\rangle &=&\frac{1}{\sqrt{2\Omega }}\sum_{j=1}^{2N_{B}}\left(
-1\right) ^{j}e^{-\alpha ^{2}\left( j-N_{\tau }\right) ^{2}}  \notag \\
&&\times e^{i\frac{\pi }{4}j}c_{B_{p},j}^{\dag }\left\vert Vac\right\rangle
\end{eqnarray}%
with the center $N_{\tau }\in \left[ 1,2N_{B}\right] $. Here we ignore the
same overall phase $e^{i\alpha ^{+}\left( \pi \right) }$ of the two cloned
GWPs. From the Eq. (\ref{ABO_GW}), we can see that the adiabatic change of
the flux leads to a relative $\pi $ phase between the two cloned GWPs.
Straightforward algebra shows that there are sinh$\left( \xi _{+}\right) $
(cosh$\left( \xi _{+}\right) $) probability on the virtual chain $a$ ($b$).
This indicates that the partial probability of wavepacket is confined in the
scatter which is different from the first case. Note that one can modulate
the value of $\Delta \delta $ to reduce the transmission probability and
therefore realize the approximate perfect confinement. In Fig. \ref{fig5}%
(b), we compute the time evolution of the GWP, which is inaccordance with
our theoretical prediction.

\section{Summary}

\label{Summary}In summary, we have systematically investigated the topology
of non-Hermitian bipartite system, the non-Hermiticity of which stems from
the staggered on-site imaginary potential. It is shown that the real part of
the Zak phase is the same with that of its Hermitian version. The existence
of the staggered imaginary potential does only effect on the imaginary part
of the Zak phase. We apply this property to a 1D non-Hermitian SSH ring
which is driven by a time-dependent magnetic flux. In the absence of the
staggered imaginary potential, the difference of Zak phase in the region of $%
\delta >0$ and $\delta <0$ is topological invariant. Therefore, the real
part of Zak phase of the concerned non-Hermitian model is topology. We can
mimic this feature through the adiabatical variation of the magnetic flux
based on the fact that the geometric phase induced by the magnetic flux is
equivalent to the Zak phase. Such topological nature of the Zak phase
motivates us to investigate a scattering problem for a time-dependent
scattering center. We find that the GWP can display two distinct dynamical
behaviors, perfect transmission or dynamical confinement, which is
determined by the timing of a flux impulse threading the ring. Specially
speaking, when the flux is added before the GWP enters into the ring, the
GWP pass perfectly through the scattering center. The GWP is confined in the
scatter partially when the flux is added during a wavepacket travelling
within the ring. Our finding provides the promising possibilities in
application of geometric phase in a non-Hermitian topological lattice system.

\section{Appendix}

\label{Sec_Appendix}

\subsection{the reduction of the scattering system and the corresponding
dynamics}

%==========================================================

\begin{figure}[tbp]
\centering
%\centering
%\includegraphics[ bb=20 100 575 782, width=0.45\textwidth, clip]{Fig4.eps}
\includegraphics[height=11.1cm,width=9cm]{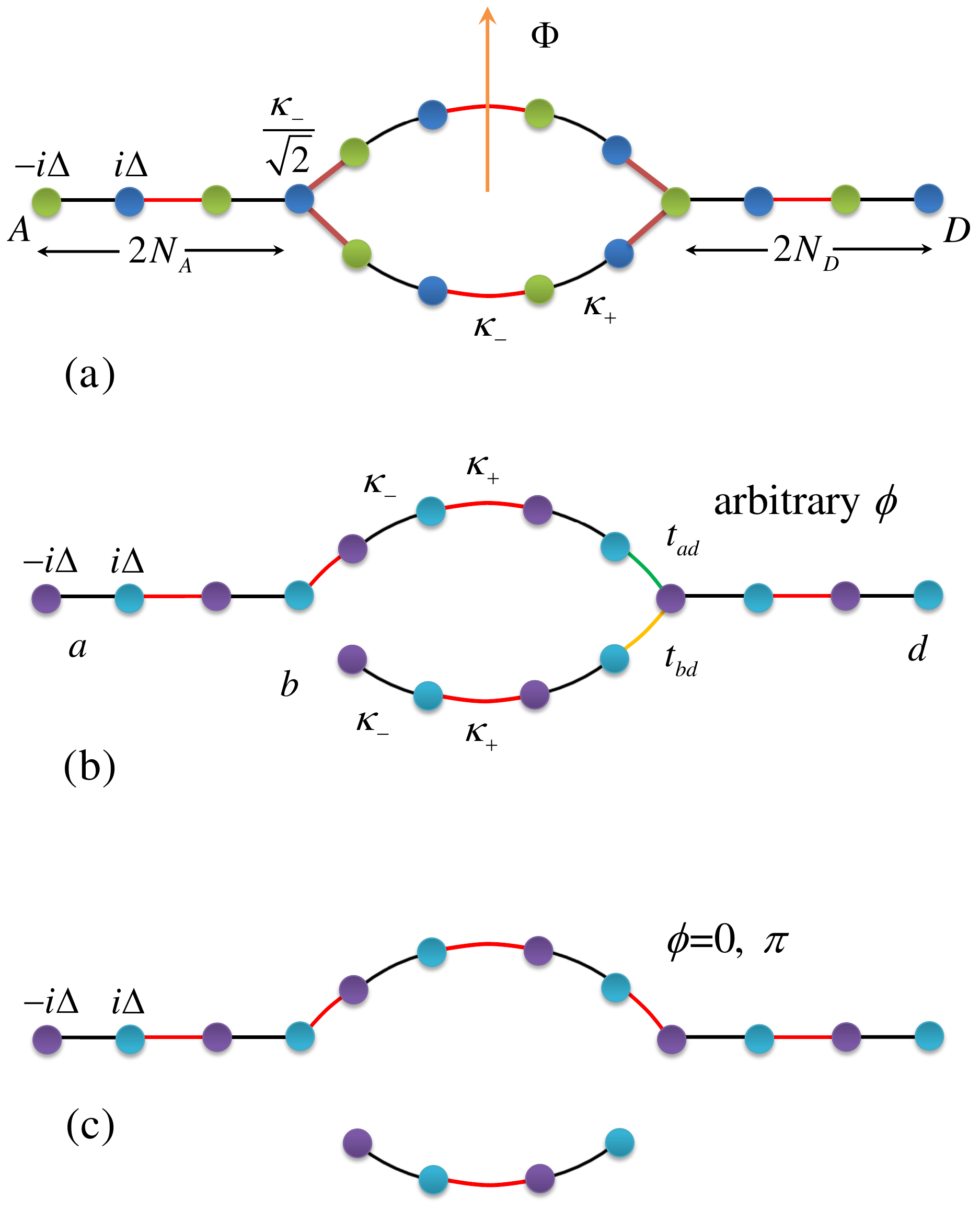}
\caption{(Color online) The $\protect\phi $-shaped scattering network with
an input non-Hermitian SSH chain $A$, an output chain non-Hermitian chain D
and a ring $\{B_{1}$, $B_{2}\}$ threaded by a magnetic flux $\Phi \left(
t\right) =4\left( N_{B}+1/2\right) \protect\phi \left( t\right) $. (b) For
an arbitrary flux, the network can be decoupled into three virtual SSH
chains $a$, $b$ and $d$. They connect with each other by the hopping
integrals $t_{ad}=\left( 1+\protect\delta \right) \cos \left[ \left(
2N_{b}+1\right) \protect\phi \right] /2$ and $t_{bd}=-i\left( 1+\protect%
\delta \right) \sin \left[ \left( 2N_{b}+1\right) \protect\phi \right] /2$.
(c) When $\protect\phi =0$, $\protect\pi $, the $\protect\phi $-shaped
scattering network can be decoupled into a long virtual non-Hermitian SSH
chain $a$ with length $N=2\left( N_{a}+N_{d}\right) $ and a short virtual
non-Hermitian SSH chain $b$ with length $2N_{b}$.}
\label{fig4}
\end{figure}

%================================================================
To reduce the network of interferometer, the four sets of new fermion
operator%
\begin{eqnarray}
\widetilde{c}_{a,l}^{\dag } &=&c_{A,l}^{\dag },  \label{t_1} \\
\widetilde{c}_{a,j+2N_{A}}^{\dag } &=&\frac{1}{\sqrt{2}}\left( e^{i\phi
j}c_{B_{1},j}^{\dag }+e^{-i\phi j}c_{B_{2},j}^{\dag }\right) ,  \label{t_2}
\\
\widetilde{c}_{b,j}^{\dag } &=&\frac{1}{\sqrt{2}}\left( e^{i\phi
j}c_{B_{1},j}^{\dag }-e^{-i\phi j}c_{B_{2},j}^{\dag }\right) ,  \label{t_3}
\\
\widetilde{c}_{d,m}^{\dag } &=&c_{D,m}^{\dag },  \label{t_4}
\end{eqnarray}%
for $l\in \left[ 1,2N_{A}\right] $, $j\in \left[ 1,2N_{B}\right] $ and $m\in %
\left[ 1,2N_{D}\right] $ are introduced to satisfy%
\begin{equation}
\left\{ \widetilde{c}_{a,j+2N_{A}}^{\dag },\text{ }\widetilde{c}%
_{b,j}\right\} =0.
\end{equation}%
The inverse transformation of the above Eqs. (\ref{t_1})-(\ref{t_4}) reduces
the Hamiltonian (\ref{H_S}) into%
\begin{equation}
H_{\mathrm{net}}=\sum_{s}\widetilde{H}_{s}+\widetilde{H}_{\mathrm{joint}},
\notag
\end{equation}%
where
\begin{eqnarray}
\widetilde{H}_{s} &=&-\frac{1}{2}\sum_{j=1}^{2N_{s}-1}\left[ 1+\left(
-1\right) ^{j}\delta \right] \left( \widetilde{c}_{s,j}^{\dag }\widetilde{c}%
_{s,j+1}+\text{\textrm{H.c.}}\right)   \notag \\
&&+i\Delta \sum_{j=1}^{2N_{s}-1}\left( -1\right) ^{j}\widetilde{c}%
_{s,j}^{\dag }\widetilde{c}_{s,j}, \\
\widetilde{H}_{\mathrm{joint}} &=&-t_{ad}\widetilde{c}_{a,2N_{a}}^{\dag }%
\widetilde{c}_{d,1}-t_{bd}\widetilde{c}_{b,2N_{b}}^{\dag }\widetilde{c}%
_{d,1}+\text{\textrm{H.c.}},
\end{eqnarray}%
with $s=a$, $b$, and $d$. The cuplings are $t_{ad}=\left( 1+\delta \right)
\cos \left[ \left( 2N_{B}+1\right) \phi \right] /2$ and $t_{bd}=iJ\left(
1+\delta \right) \sin \left[ \left( 2N_{B}+1\right) \phi \right] /2$,
respectively. For clarity, we sketch this decomposition in Fig.\ref{fig4}.
It is shown that for an arbitrary flux $\phi $, the concerned network can be
decoupled into three virtual non-Hermitian SSH chains $a$, $b$ and $d$ with
length $N_{a}=N_{A}+N_{B}$, $N_{b}=N_{B}$ and $N_{d}=N_{d}$, respectively.
The virtual chains $a$ and $b$ connect with chain $d$ through the hopping
integral $t_{ad}$ and $t_{bd}$, which depends on the magnetic flux $\phi $.
In the following, we focus on the case that $\phi =0$ or $\pi $. Under this
condition, the virtual chains $b$ and $d$ are decoupled. The network reduced
to two independent non-Hermitian SSH chains with length $N=2N_{a}+2N_{d}$
and $2N_{b}$, respectively. Then the corresponding Hamiltonian can be given
as
\begin{eqnarray}
H_{\mathrm{net}} &=&-\frac{1}{2}\sum_{j=1}^{2N-1}\left[ 1+\left( -1\right)
^{j}\delta \right] \left( \widetilde{c}_{a,j}^{\dag }\widetilde{c}_{a,j+1}+%
\text{\textrm{H.c.}}\right)   \notag \\
&&+i\Delta \sum_{j=1}^{2N-1}\left( -1\right) ^{j}\widetilde{c}_{a,j}^{\dag }%
\widetilde{c}_{a,j},  \notag \\
&&-\frac{1}{2}\sum_{j=1}^{2N_{b}-1}\left[ 1+\left( -1\right) ^{j}\delta %
\right] \left( \widetilde{c}_{b,j}^{\dag }\widetilde{c}_{b,j+1}+\text{%
\textrm{H.c.}}\right)   \notag \\
&&+i\Delta \sum_{j=1}^{2N_{b}-1}\left( -1\right) ^{j}\widetilde{c}%
_{b,j}^{\dag }\widetilde{c}_{b,j},
\end{eqnarray}%
with newly defined operators
\begin{equation}
\widetilde{c}_{a,2N_{A}+2N_{B}+m}^{\dag }=\widetilde{c}_{d,m}^{\dag }.
\end{equation}%
This fact means that for an arbitrary initial state localized on the virtual
chain $a\left( b\right) $, it will evolve driven by the virtual chain of
length $N\left( 2N_{b}\right) $. There are two typical features that should
be mentioned: (i) For the state localized on the virtual chain $a$, the
evolve state will exhibit perfect transmission through the scattering
center. (ii) For the state localized on the virtual chain $b$, the localized
state will be confined in the scattering center. These two mechanisms
are crucial to understand the wavepacket dynamics.

\acknowledgments This work was supported by National Natural Science
Foundation of China (under Grants No. 11505126, No. 11874225). X.Z.Z. was
also supported by the Ph.D. research startup foundation of Tianjin Normal
University under Grant No. 52XB1415, and the Program for Innovative Research
in University of Tianjin (under Grant No. TD13-5077).

\end{document}